\documentclass[a4paper]{jpconf}
\usepackage[english]{babel}
\usepackage[utf8]{inputenc}
\usepackage{amsthm}
\usepackage{mathtools}
\usepackage{physics}
\usepackage[colorinlistoftodos, color=green!40, prependcaption]{todonotes}
\usepackage{dsfont}
\usepackage{chemformula}
\usepackage{units}
\usepackage{amsmath}
\usepackage{relsize}
\usepackage{enumitem}
\usepackage{epsfig}
\usepackage[amssymb]{SIunits}
\usepackage[square,sort&compress,numbers]{natbib}
\usepackage{acronym}

\newacro{mc}[MC]{Monte Carlo}
\newacro{cpa}[CPA]{coherent potential approximation}
\newacro{rpa}[RPA]{random phase approximation}
\newacro{bz}[BZ]{Brillouin zone}

\begin{document}
\title{Eigenmodes of a disordered FeCo magnonic crystal at finite temperatures} 
\author{S Paischer$^1$, P A Buczek$^2$, N Buczek$^3$, D Eilmsteiner$^{1,2}$ and A Ernst$^{1,4}$}
\address{$^1$Institute for  Theoretical Physics, Johannes Kepler  University Linz, Altenberger  Stra{\ss}e 69, 4040 Linz}
\address{$^2$Department of Engineering and Computer Sciences, Hamburg  University of Applied Sciences, Berliner Tor 7, 20099 Hamburg, Germany}
\address{$^3$Department of Applied Natural Sciences, L\"ubeck  University of Applied Sciences, M\"onkhofer Weg 239, 23562 L\"ubeck,  Germany}
\address{$^4$Max-Planck-Institut of Microstructure Physics, Weinberg  2, 06120 Halle (Saale), Germany}
\ead{sebastian.paischer@jku.at}

\begin{abstract}
  In this report we present a systematic study of the magnonic modes in the disordered \ch{Fe_{0.5}Co_{0.5}} alloy based on the Heisenberg Hamiltonian using two complementary approaches. In order to account for substitutional disorder, on the one hand we directly average the transverse magnetic susceptibility in real space over different disorder configurations and on the other hand we use the coherent potential approximation (CPA). While the method of direct averaging is numerically exact, it is computationally expensive and limited by the maximal size of the supercell which can be simulated on a computer. On the contrary the CPA does not suffer from this drawback and yields a cheap numerical scheme. Therefore, we additionally compare the results of these two approaches and show that the CPA gives very good results for most of the magnetic properties considered in this report, including the magnon energies and the spatial shape of the eigenmodes. However, it turns out that while reproducing the general trend, the CPA systematically underestimates the disorder induced damping of the magnons. This provides evidence that the physics of impurity scattering in this system is governed by non-local effects missing in the CPA. Finally, we study the real space eigenmodes of the system, including their spatial shapes, and analyze their temperature dependence within the random phase approximation.
\end{abstract}



\section{Introduction}

Over the last few decades, the field of magnon spintronics, or
magnonics, gained an ever increasing amount of attention. This novel
strategy of data propagation and processing has several advantages
over the commonly used electronic circuits, like a lack of energy
loss through Joule heating \cite{Chumak2015}. The elemental information
carriers are spin waves (also called magnons) which can be pictured
as a coherent precession of the magnetic moments in the material
\cite{Bloch1930}. Similar to phonons, magnons are Bloch waves in
periodic systems carrying a crystal momentum and energy. However, in
order to construct a magnonic circuit, one depends on suitable
materials, referred to as magnonic crystals
\cite{Chumak2017,Nikitov2001}. The most common magnetic materials are
not suited for the use in magnonics since they lack desired properties,
especially the emergence of a magnonic bandgap (i.e. frequency bands
in which magnon states cannot propagate in the solid
\cite{Krawczyk2014,Lenk2011}). Combined with the unique spin wave
dispersion close to the band edges, this feature provides a rich
toolbox for magnon mode engineering, including the possibility of
selective spin wave excitations and propagation, magnon mode
confinement and deceleration, and bandgap soliton generation \cite{Sadovnikov2016,Sadovnikov2018,Sheshukova2013}.\\\\


In current research, mostly long wavelength magnons with energies in
the gigahertz band are studied. However, in principle magnons in the
terrahertz regime are preferred for magnonic applications as they
warrant faster information transport and smaller devices
\cite{Zakeri2018}. In materials with many magnetic atoms in the
primitive unit cell, one expects the occurrence of several magnon
modes, which might be separated by a bandgap, yielding a natural
magnonic crystal \cite{Buczek2009}.


Here, we concentrate on the ferromagnetic \ch{Fe_{0.5}Co_{0.5}} alloy.
This alloy shows all the necessary properties for a terahertz magnonic
crystal: Its typical magnon energies are well within the terahertz
range, it has a high Curie temperature
\cite{Normanton1975,Nishizawa1984}, and \ac{cpa} studies suggest the
spectrum to exhibit a bandgap
which remains stable at elevated temperatures \cite{Paischer2021}.  It
is interesting to note that magnonic crystals used in gigahertz
applications are typically artificial heterostructures obtained from
elaborate fabrication processes \cite{Chumak2017}.  On the contrary,
in the terahertz range, the natural microscopic arrangement of atoms
in alloys like \ch{Fe_{x}Co_{1-x}} would suffice to create cheap
magnonic crystals.

In real materials, there are several mechanisms that influence the
lifetimes of magnons. First, the interaction of magnons with
electronic excitations including a spin flip, called Stoner
excitations, plays an essential role especially in metals
\cite{Buczek2011,Buczek2011a,Zhang2012}. This mechanism, called Landau
damping, was shown
to be affected by reduced dimensionality of the
system and alloying \cite{Qin2015}. 

Second, the scattering on crystal imperfections might influence the
magnon lifetimes as well. In our recent report \cite{Paischer2021} we showed
that this effect may lead to non-trivial dependence of the magnon
damping in iron cobalt alloys when the concentration of cobalt is
varied. 
Finally, a non zero temperature is expected to reduce the lifetimes
of magnons in the system. 
Materials to be used for magnonic devices have to operate well above
room temperature and feature structural imperfections, as every solid does. Thus, it is interesting to investigate how the magnonic properties evolve in real,
imperfect or alloyed solids at non-zero temperatures. We analyze the
alloy \ch{Fe_{0.5}Co_{0.5}} using two different approaches: On the one
hand, we directly average the transverse magnetic susceptibility over
several disordered configurations. Only substitutional disorder is
considered which is generated using pseudorandom numbers. Therefore
we resort to this method as \ac{mc} method in the following. On the
other hand we utilize a \ac{cpa} applied to the disordered
Heisenberg ferromagnet \cite{Buczek2016}. This mean field approach was successfully applied for the calculation of electronic and magnetic properties in numerous materials, e.g. \cite{Buczek2016,Edstroem2015,Rusz2006}. The superiority of our
\ac{cpa} method compared to other treatments of the same problem is
the possibility to account for complex crystal structures. To
incorporate finite temperature effects, we implemented a modified
version of the \ac{rpa} discussed in reference \cite{Callen1963}. While the \ac{rpa} method accounts quantitatively for the softening of the magnon modes with temperature, it does not describe the reduction of the magnon life-time due to the interaction of these modes with the thermal bath mentioned above. We
show that both \ac{mc} and \ac{cpa} methods give the same magnetic spectrum and the spatial
shape of the eigenmodes. The only discrepancy between the two methods
appears in the magnitude of disorder-induced damping which is clearly underestimated within
the \ac{cpa}. Furthermore, we analyze the real space representation of
the dominant eigenmode at 200~meV and show that, with the \ac{rpa}, the spatial shape of the modes are basically
unaltered by increasing temperatures.  Our formalism
does not include the Landau damping of the spin waves. This
attenuation mechanism can be pronounced in metallic magnonic crystals
and can be described within the framework of many-body perturbation
theory \cite{Bluegel2013} or time-dependent density functional theory
\cite{Staunton1999,Staunton2000,Buczek2011a}. Although the dynamical
magnetic susceptibility can be calculated for disordered materials
within a \ac{cpa} method \cite{Staunton1999,Staunton2000}, the approach requires careful numerical analysis and is subject of a separate study.     
%

The paper is organized as follows: In chapter \ref{chap_theory}, the
theoretical background of the \ac{rpa}-\ac{cpa} theory and the \ac{mc}
method for the disordered Heisenberg ferromagnet are discussed. The
obtained results are presented in chapter \ref{chap_res}.

\section{Theory}\label{chap_theory}
We deploy the following form of the Heisenberg Hamiltonian
\begin{align}
  H=-\frac{1}{2}\sum_{i,j}J_{ij}~\vb*{e}_i\cdot\vb*{e}_j\label{eqn_hh}
\end{align}
where $J_{ij}$ are the exchange parameters which were obtained from
the magnetic force theorem \citep{Liechtenstein1987} and $\vb*{e}_i$ is
a unit vector in the direction of the magnetization at site $i$. Anisotropy terms are neglected on the energy scales relevant for this study. To calculate
magnon properties, the transverse magnetic susceptibility \citep{Nolting2009}
\begin{align}
  \chi_{ij}(t,t')=-\text{i}~\Theta(t-t')~\overline{\comm{\mu_i^+(t)}{\mu_j^-(t')}}\label{eqn_susc_def}
\end{align}
with $\mu_i^\pm=\mu_i^x\pm\text{i}\mu_i^y$, $\mu_i^\alpha$ being the
$\alpha$-component of the magnetic moment $\vb*{\mu}_i$ on the lattice 
site $i$ and the overline represents a thermal average, is
computed. It can be found by solving the equation of motion 
\begin{align}
z\chi_{ij}(z)=2g\delta_{ij}~\overline{\mu}_i-g\sum_{\ell}\frac{\overline{\mu}_i}{\mu_i\mu_\ell} J_{i\ell}~\chi_{\ell j}(z)+g\sum_{\ell}\frac{\overline{\mu }_\ell}{\mu_i\mu_\ell}J_{i\ell}~\chi_{ij}(z)~.\label{eqn_susc}
\end{align}
with the energy $z=E+\text{i}\epsilon$ and the Landé factor $g$.
In the following, we assume a material with a complex structure and use
an argument $\vb*{R}$ to specify the primitive unit cell, a latin
index to specify the basis site and a greek index to distinguish
between different atomic species. 
The disorder is modeled by defining
occupation variables
\begin{align}
  p^{i\alpha}(\vb*{R})=\left\{\begin{array}{lr}
      1&\text{species $\alpha$ on basis site $i$ in unit cell $\vb*{R}$}\\
      0&\text{else}
    \end{array}\right.
\end{align}
and a species resolved Fourier transformation
\begin{align}
  \chi^{\alpha\beta}_{ij}(z,\vb*{k},\vb*{k'}):=\sum_{\vb*{R},\vb*{R}'}p^{i\alpha}(\vb*{R})~\text{e}^{-\text{i}\vb*{k}\cdot\vb*{R}}~\chi_{ij}(\vb*{R},\vb*{R'})~p^{j\beta}(\vb*{R'})~\text{e}^{\text{i}\vb*{k'}\cdot\vb*{R'}}.\label{eqn_Ft}
\end{align}
We calculate this species resolved susceptibility and average it using
the \ac{mc} and \ac{cpa} methods described in more detail later in this
section. 

The averaged susceptibility $\vb*{\mathcal{X}}$ is used to calculate the loss matrix
\begin{align}\label{eqn_loss}
\vb*{\mathcal{L}}(z,\vb*{k})=\frac{1}{2\text{i}}\left(\vb*{\mathcal{X}}(z,\vb*{k})-\vb*{\mathcal{X}}^\dagger(z,\vb*{k})\right),
\end{align}
which holds the information about the elementary excitations of the system.
Its eigenvalues are non zero at the energies and wavevectors at which
elementary excitations, i.e. magnons, occur. The eigenvectors at the corresponding energy and wavevector yield the
shape of the excitation in question. 
The components of the eigenvectors are to be
interpreted as the transverse component of the magnetic moments induced during the moment's precession (magnon excitation). The fluctuation-dissipation theorem \cite{Kubo1966} states that the excited states of the system are intrinsically related to the linear response of the system upon application of the corresponding external perturbation. Therefore, the shape of magnonic modes can be inferred from the analysis of the response of the magnet to an external magnetic field, in this case in the direction perpendicular to the ground state magnetization.

\begin{figure}
	\centering
	\includegraphics[width=4cm]{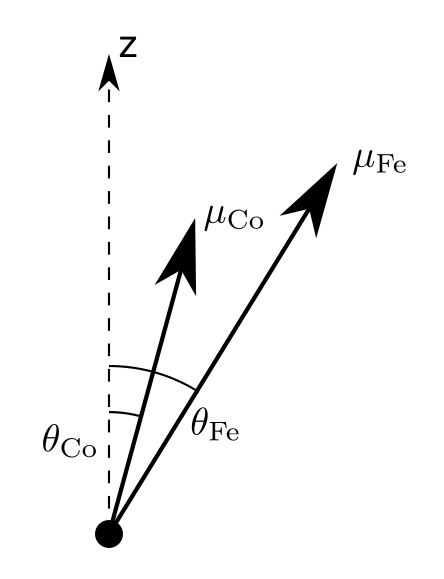}
	\caption{Angles in the case of the iron cobalt alloy.}
	\label{fig_angles}
\end{figure}

In the linear regime, the small angle $\theta$ between the magnetic moments and the $z$ axis (giving the direction of the ground state magnetization)
depends on the strength of the external perturbation, cf. figure
\ref{fig_angles}. It is zero in the ground state. 
Suppose the projection of the tilted moments to the $xy$ plane be
$\epsilon\tilde{\mu}_{i\alpha}$ with a small parameter $\epsilon$
linearly dependent on the strength of the external field. The component of
the (normalized) eigenvector of the loss matrix is
$\tilde{\mu}_{i\alpha}$. Then
\begin{align}
\theta_{i\alpha}\approx\sin(\theta_{i\alpha})=\frac{\epsilon\tilde{\mu}_{i\alpha}^{xy}}{\mu_{i\alpha}}.
\end{align}
The ratio between the angles $\theta_{i\alpha}$ of different
constituents is independent of $\epsilon$.
\begin{align}
\frac{\theta_{i\alpha}}{\theta_{j\beta}}=\frac{\tilde{\mu}_{i\alpha}\mu_{j\beta}}{\tilde{\mu}_{j\beta}\mu_{i\alpha}}\label{eqn_theta}
\end{align}

We model the temperature dependence by a generalized version of the
\ac{rpa} introduced by \emph{Callen} \cite{Callen1963} for simple
ordered systems. The thermally averaged magnetic moments are evaluated
by calculating 
\begin{align}
	\overline{\mu}=g\frac{\left(\frac{\mu}{g}-\Phi\right)(1+\Phi)^{\mu+1}+\left(\frac{\mu}{g}+1+\Phi\right)\Phi^{\mu+1}}{\left(1+\Phi\right)^{\mu+1}-\Phi^{\mu+1}}
\end{align}
once for every species for \ac{cpa}, and once for every site in the system for \ac{mc} calculations. Here,
\begin{align}
	\Phi=\mathop{\mathlarger{\mathlarger{\int}}}_{-\infty}^{\infty}\dd{z}\frac{D(z)}{\text{e}^{\frac{z}{k_{\text{B}}T}}-1}\label{eqn_phi}
\end{align}
with the density of states $D(z)$ for the species or lattice site in question.
\subsection{\ac{mc} calculations}
For the direct numerical averaging of the susceptibility, we use a
supercell with 20$\times$20$\times$20 primitive unit cells and
periodic boundary conditions. The generation of the random occupation
is done by means of pseudorandom numbers. 
The obtained real space susceptibility is averaged over 30 different
configurations. With these parameters, the results are converged. They will be presented in chapter \ref{chap_res}. 
The susceptibility of each configuration is given in real space by
\begin{align}
	\vb*{\chi}=\vb*{M}^{-1}\vb*{\mu}
\end{align}
where bold symbols denote matrices in the site basis and
\begin{align}
\mu_{ij}&=2g\delta_{ij}~\overline{\mu}_i\nonumber\\
M_{ij}&=z\delta_{ij}+g\frac{\overline{\mu}_i}{\mu_i\mu_j} J_{ij}-g\sum_m\frac{\overline{\mu }_m}{\mu_i\mu_m}J_{im}\delta_{ij}~.
\end{align}

\subsection{Coherent Potential Approximation}
\begin{figure}
	\includegraphics[width=8cm]{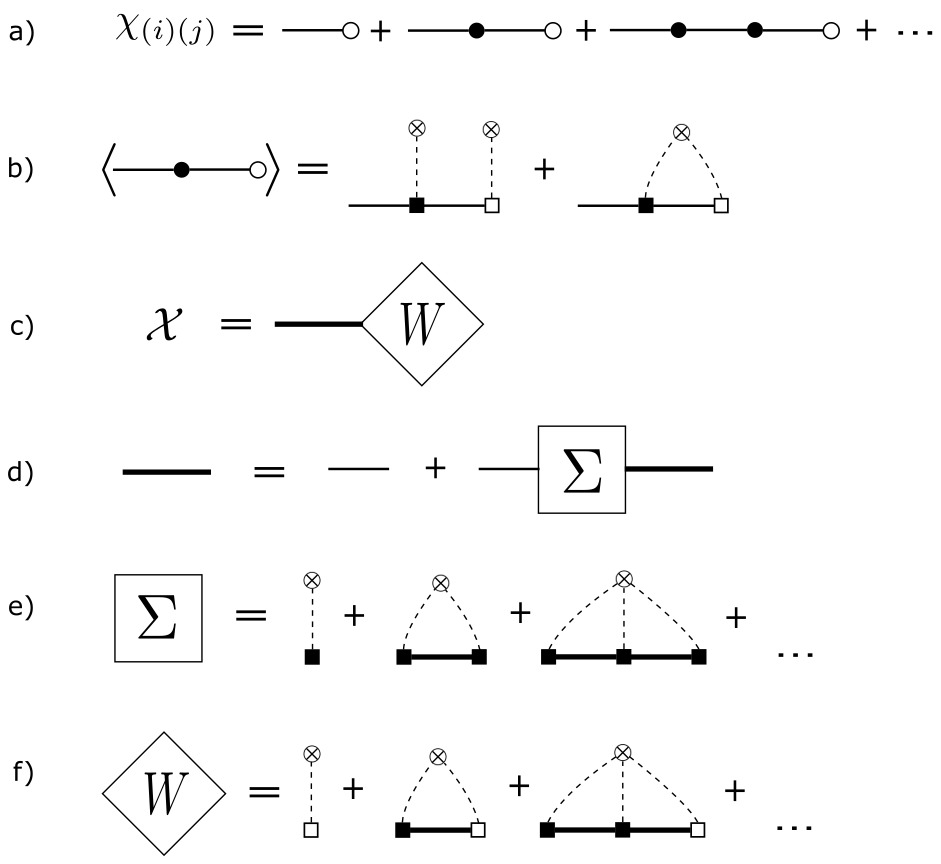}
	\centering
	\caption{Diagrammatic representation of the main results of the
		\ac{cpa}-theory. a) Fourier transformation of series \eqref{eqn_susc},
		b) average of the second term in a), c) the averaged
		susceptibility $\mathcal{X}$ written as a product of the effective
		medium propagator $\varXi$ (thick line) and the spin weight $W$, d)
		Dyson equation for the effective medium propagator, e) definition
		of the self-energy $\varSigma$ and f) definition of the spin
		weight $W$}
	\label{fig_feyn}
\end{figure}
Within the \ac{cpa} formalism, the equation of motion \ref{eqn_susc}
is used to generate a series expansion of the susceptibility. The
series is the Fourier transformed using equation \ref{eqn_Ft}, which
leads to the series written diagrammatically in figure \ref{fig_feyn}
a) \cite{Yonezawa1968}. The following symbols are used:
\begin{itemize}
\item The $\tau$-matrix
  \begin{align}
    \tau^{(\ell)}_{(i)(j)}(\vb*{k},\vb*{k}')=g\mu_{(j)}^{-1}&\left(J_{(j)(\ell)}(\vb*{k}-\vb*{k'})\frac{\overline{\mu}_{(\ell)}}{\mu_{(\ell)}}\delta_{(i)(j)}\right.\nonumber\\&\left.-J_{(\ell)(j)}(\vb*{k}')\frac{\overline{\mu}_{(i)}}{\mu_{(i)}}\delta_{(i)(\ell)}\right) 
  \end{align}
  where
  \begin{align}
    J_{(i)(j)}(\vb*{k})&=\sum_{\vb*{R}}J_{(i)(j)}(\vb*{R})~\text{e}^{-\text{i}\vb*{k}\cdot\vb*{R}}
  \end{align}
  is represented by a filled square.
\item The filled circle represents a $T$- matrix
  \begin{align}
    T_{(i)(j)}(\vb*{k},\vb*{k}')=\sum_{(\ell)}\varrho^{(\ell)}(\vb*{k}-\vb*{k'})~\tau^{(\ell)}_{(i)(j)}(\vb*{k},\vb*{k}').
  \end{align}
\item An empty square stands for a $\sigma$-matrix:
  \begin{align}
    \sigma^{(\ell)}_{(i)(j)}=2g\delta_{(i)(j)}\delta_{(i)(\ell)}\overline{\mu}_{(\ell)}
  \end{align}
\item The $S$-matrix is depicted as an empty circle and is given by
  \begin{align}
    S_{(i)(j)}(\vb*{k},\vb*{k'})=\sum_{(\ell)}\varrho^{(\ell)}(\vb*{k}-\vb*{k'})~\sigma^{(\ell)}_{(i)(j)}.
  \end{align}
\item The propagator of uncoupled magnetic moments, represented by a solid line, is given as
  \begin{align}
    \varGamma_{(i)(j)}(z)=z^{-1}\delta_{(i)(j)}.
  \end{align}
\item A cumulant of order $n$ is represented by a crossed
  circle, where the order is given by the number of dashed lines
  ending at it.
\end{itemize}
Furthermore, two rules for the interpretation of the diagrams need to
be followed:
\begin{enumerate}
\item The elements brought together in a diagram undergo a matrix
  multiplication in the $(i)(j)$-space. The corresponding matrix
  indices are written as subscripts in the definitions above.
\item Every internal free propagator is assigned a momentum which is
  integrated over:
  \begin{align}
    \frac{1}{\varOmega_{\text{BZ}}}\int_{\varOmega_{\text{BZ}}}\dd[3]{k_1}
  \end{align}
\end{enumerate}
Averaging over all possible configurations needs to be done
carefully to obtain correct magnetic properties of the
material. The average of the Fourier transformed second term in series 
\ref{eqn_susc} is depicted in figure \ref{fig_feyn} b). Starting from
the fourth order diagrams, diagrams with
crossed dashed lines will appear. These diagrams correspond to
correlations between the occupation of different sites. In our
approach, these diagrams are neglected. Since the averaged diagrams
consist of two different types of vertices (filled and empty squares),
the averaged susceptibility $\vb*{\mathcal{X}}$ can be written as a
product of two quantities as shown in figure \ref{fig_feyn} c). These
two quantities are defined in \ref{fig_feyn} d) - f). It can be
shown that all non-crossed diagrams can be constructed with these
definitions. The self consistency of the method is evident from the
fact that the self energy (figure \ref{fig_feyn} e)) depends on the
effective medium propagator which in turn depends on the self
energy. Further details of the theory are given in references \cite{Paischer2021,Buczek2016,Buczek2018}. 
\section{Results}\label{chap_res}
Magnetic moments $\mu_i^\alpha$ and exchange parameters $J_{ij}$ of
iron-cobalt alloys at various concentrations were evaluated using a
first-principles Green-function method within a generalized gradient
approximation of density functional theory~\citep{Perdew1996}.  The
method is designed for bulk materials, surfaces, interfaces and real
space clusters~\citep{Luders2001,Geilhufe2015,Hoffmann2020}. The resulting magnetic moments read $\mu_\text{Fe}=2.65\mu_\text{B}$ and $\mu_\text{Co}=1.83\mu_\text{B}$. The average magnetic moment per atom is given by the arithmetic mean of the constituents magnetic moments (for 50:50 alloys) and reads $\mu=2.24\mu_\text{B}$, which is in good agreement with experimental results \cite{Goldman1949}. The impact of disorder on the electronic structure was taken into account within the electronic \ac{cpa} \citep{Soven1967}
 implemented within multiple scattering
theory~\citep{Gyorffy1972}. Exchange interaction was estimated using
the magnetic force theorem~\citep{Liechtenstein1987} formulated for
substitutional alloys within the \ac{cpa} approach~\citep{Turek2006}.
Although both iron and cobalt are known for long ranged interaction
between the magnetic moments, the results presented in this section
use only 12 shells of neighbors because of computational limits. To
ensure the convergence of spin waves with the number of neighbor
shells and supercell size, several calculations were performed for 30
neighbor shells showing essentially the same results as with 12
shells. We consider only the disordered alloy \ch{Fe_{0.5}Co_{0.5}}.

\begin{figure}
	\centering
	\includegraphics[width=0.45\textwidth]{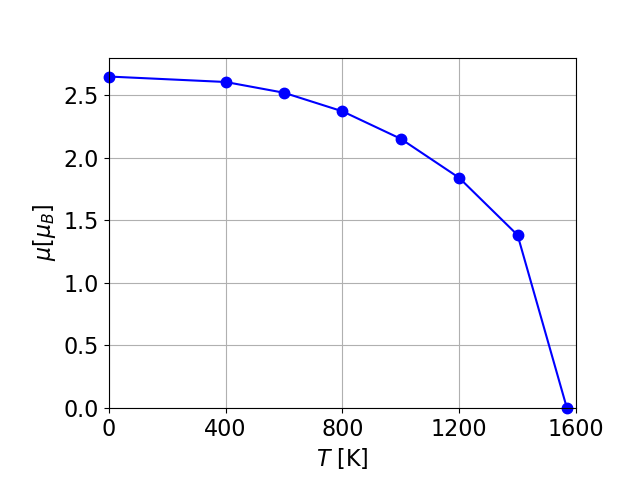}
	\caption{Calculated temperature dependence of the magnetic moments' $z$ component. The critical temperature lies at $T_\text{C}=1569$K.}
	\label{fig_tm}
\end{figure}

For better comparability the bcc structure was taken in all calculations. Furthermore, the interaction parameters $J_{ij}$ are
held constant (at their value at $T=0\text{K}$). 
Both \ac{mc} and \ac{cpa} calculations are done at
complex energies with a small imaginary part
$\epsilon=10^{-4}\text{Ry}$.\\
The resulting temperature dependence of the magnetic moments is shown in figure \ref{fig_tm}. Although the RPA is known to underestimate the Curie temperature \cite{Rusz2005}, we obtain a Curie temperature above the experimental values of $T_\text{C}\approx1250$K \cite{Normanton1975,Nishizawa1984}. This behavior can be partly explained by the fact that the real FeCo system will perform a structural phase transition, while in the calculations the structure (bcc) was fixed.
\subsection{Eigenmodes at $T=0$}
\begin{figure}
	\centering
	\includegraphics[width=0.45\textwidth]{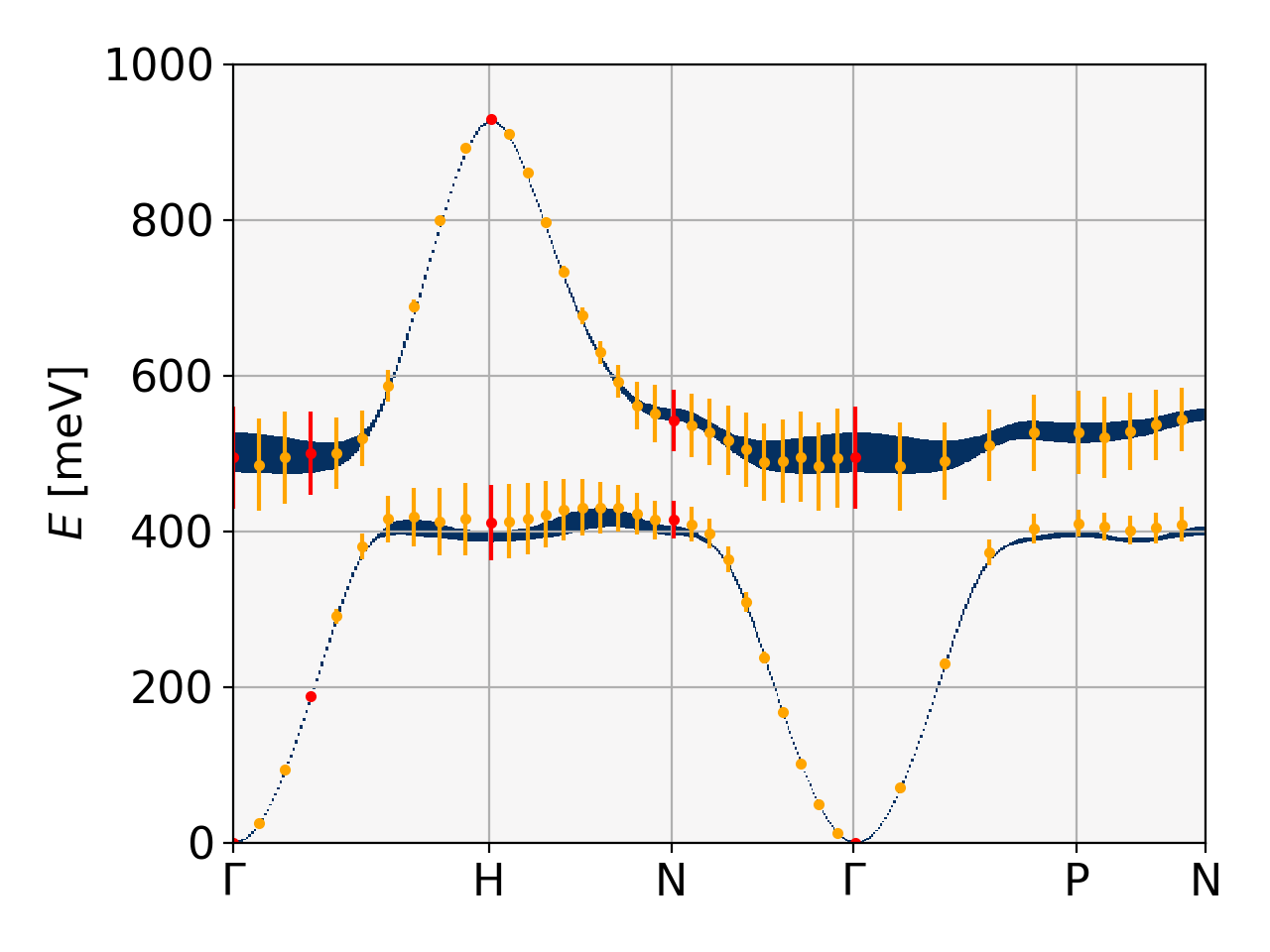}
	\includegraphics[width=0.45\textwidth]{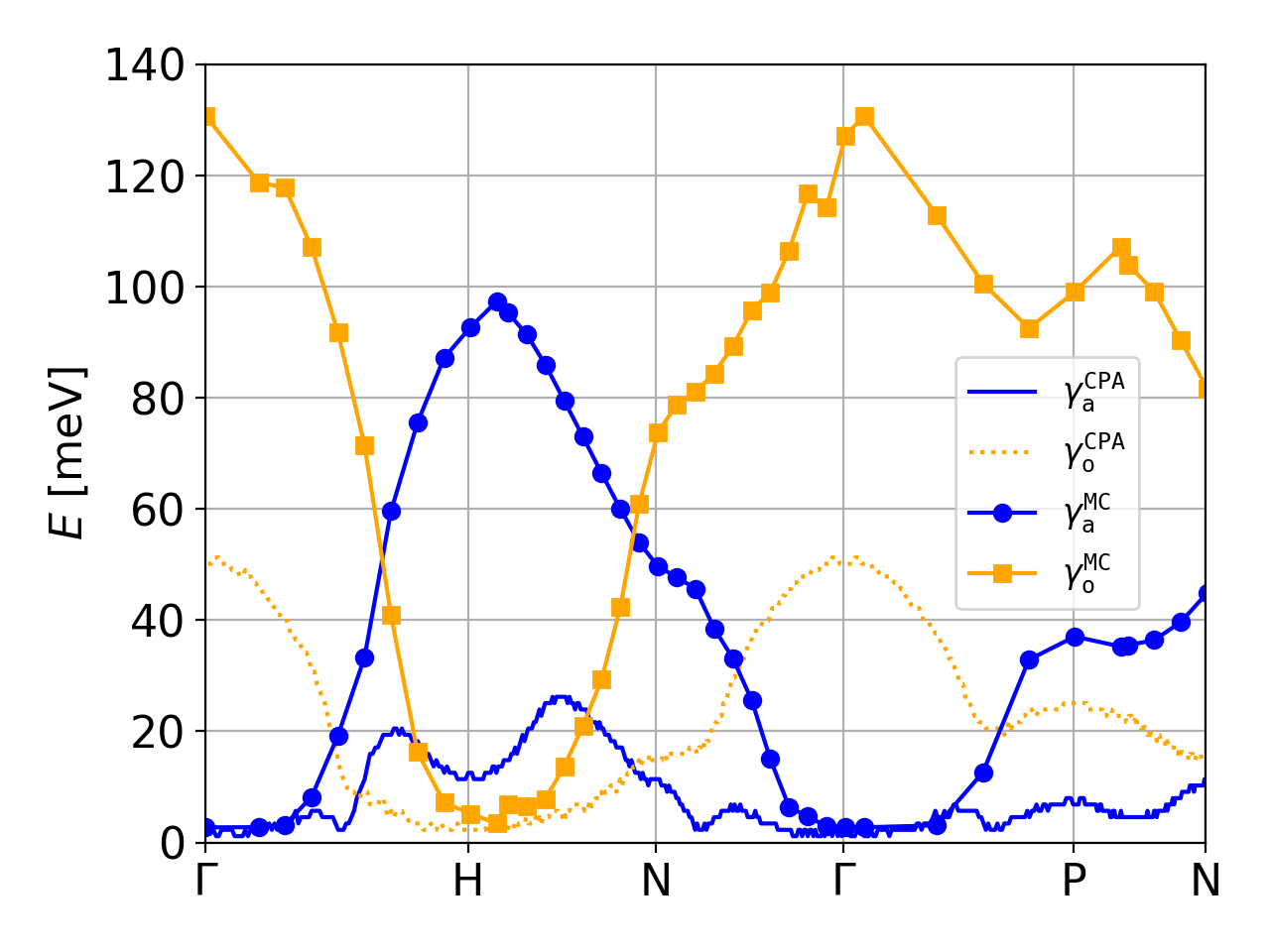}
	\caption{Comparison of the magnonic spectra (left) obtained
          through the \ac{cpa} (blue lines) and MC (orange and red
          dots) calculations. The errorbars for the MC-results
          represent the width of the peaks as does the line width for
          the \ac{cpa} results. In the right figure the absolute values
          of the peak widths are shown.} 
	\label{fig_spectrum}
\end{figure}
The magnonic spectra (cf. figure \ref{fig_spectrum}) obtained using
both methods are basically identical. In the left plot of figure
\ref{fig_spectrum}, the density plot in the background represents the
imaginary part of the \ac{cpa} susceptibility while the dots stand for
the position of the susceptibility peaks within the MC. The errorbars
at the points represent the full width at half maximum of the peaks. The same holds for the
width of the line and the \ac{cpa} results. Due to the finite size of
the supercell, the \ac{mc} can only give meaningful results for a
discrete set of points within the \ac{bz}. The spectra are in good agreement with other results for ordered FeCo systems \cite{Okumura2019,Grotheer2001} (note that the spectra in these references are calculated for a CsCl structure). \\
We obtain a spin wave stiffness of $C\approx675$meV\AA\ at $T=0$K. Experimental values range from $C\approx500$meV\AA\ \cite{Lowde1965}(at $T=0$K) to $C\approx800$meV\AA\ \cite{Liu1994}(thin film at room temperature). We note that the long range interaction between magnetic moments which was neglected in this work due to computational limits may influence the magnonic spectrum, especially close to the $\Gamma$ point and therefore also affect the spin stiffness.

It can clearly be seen that the \ac{cpa} systematically underestimates
the width of the peaks. In the right plot of figure
\ref{fig_spectrum}, we show the full width at half maximum $\gamma$
for the acoustic and optic modes within the \ac{cpa} and the \ac{mc}
calculations. Although the absolute values of the damping is
different, the general trend of $\gamma$ through the considered paths
in the \ac{bz} is very similar. The fact that the MC calculations give
much larger widths indicates that non-local effects play an essential
role in the damping of magnons. Recent studies \cite{Zhang2007,Zhang2017,Zhang2020} reveal, that non-local effects also play an important role in the Ising model. Although the Ising model only captures nearest neighbor spin interaction, in 3D a term with an effective long range interaction appears in the partition function \cite{Zhang2017}. Although the results of these studies cannot be directly applied to our model, they represent a further hint at the presence and influence of non-local effects.\\
 The CPA predicts [14] that at the 50\% concentration of cobalt the bandgap is still present. However, the more realistic estimation of the magnon damping using the MC method tends to suggest that the significant widening of the magnon modes might lead to effective closing of the gap.

\begin{figure}
  \centering
  \includegraphics[width=0.45\textwidth]{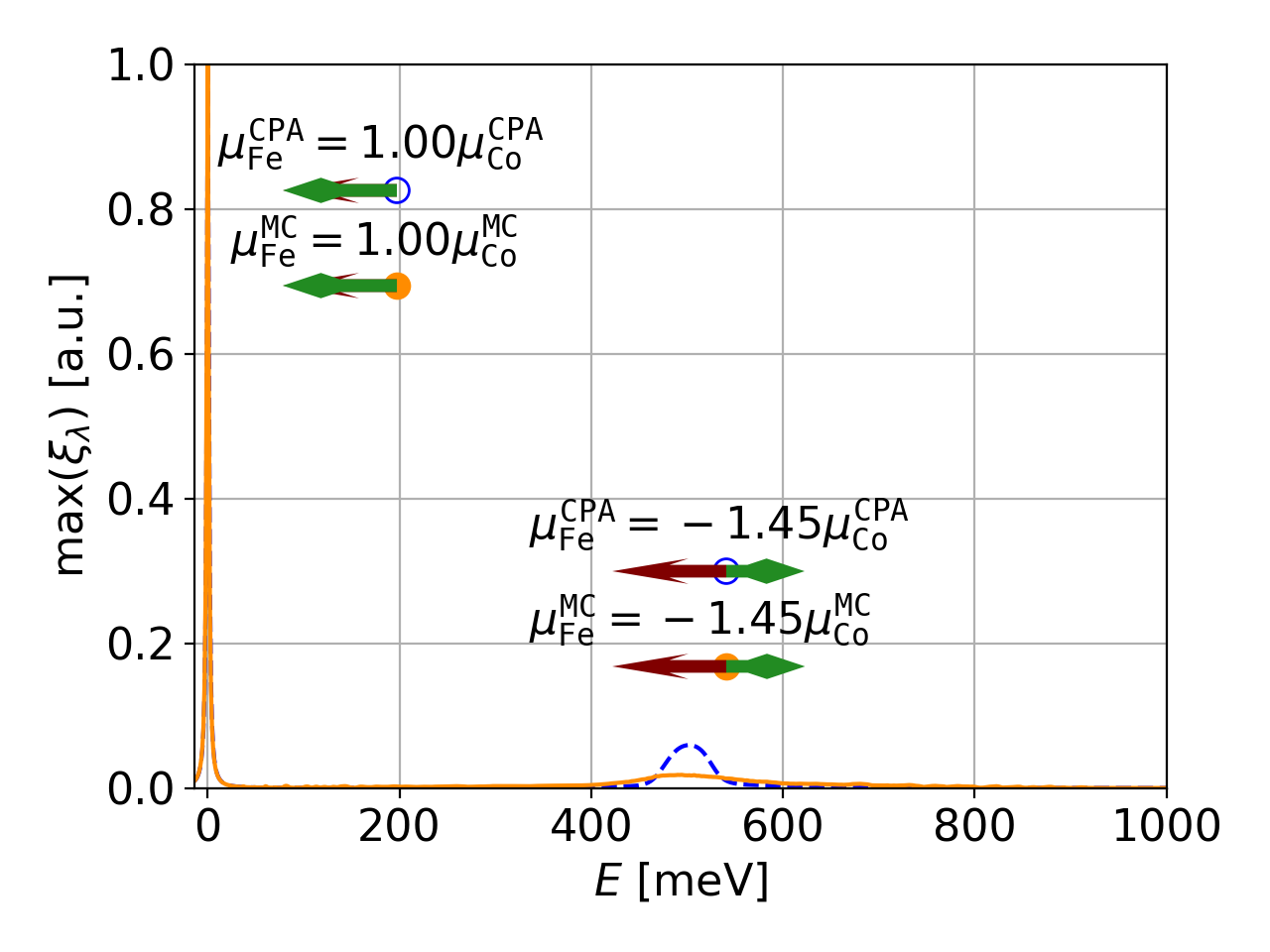}
  \includegraphics[width=0.45\textwidth]{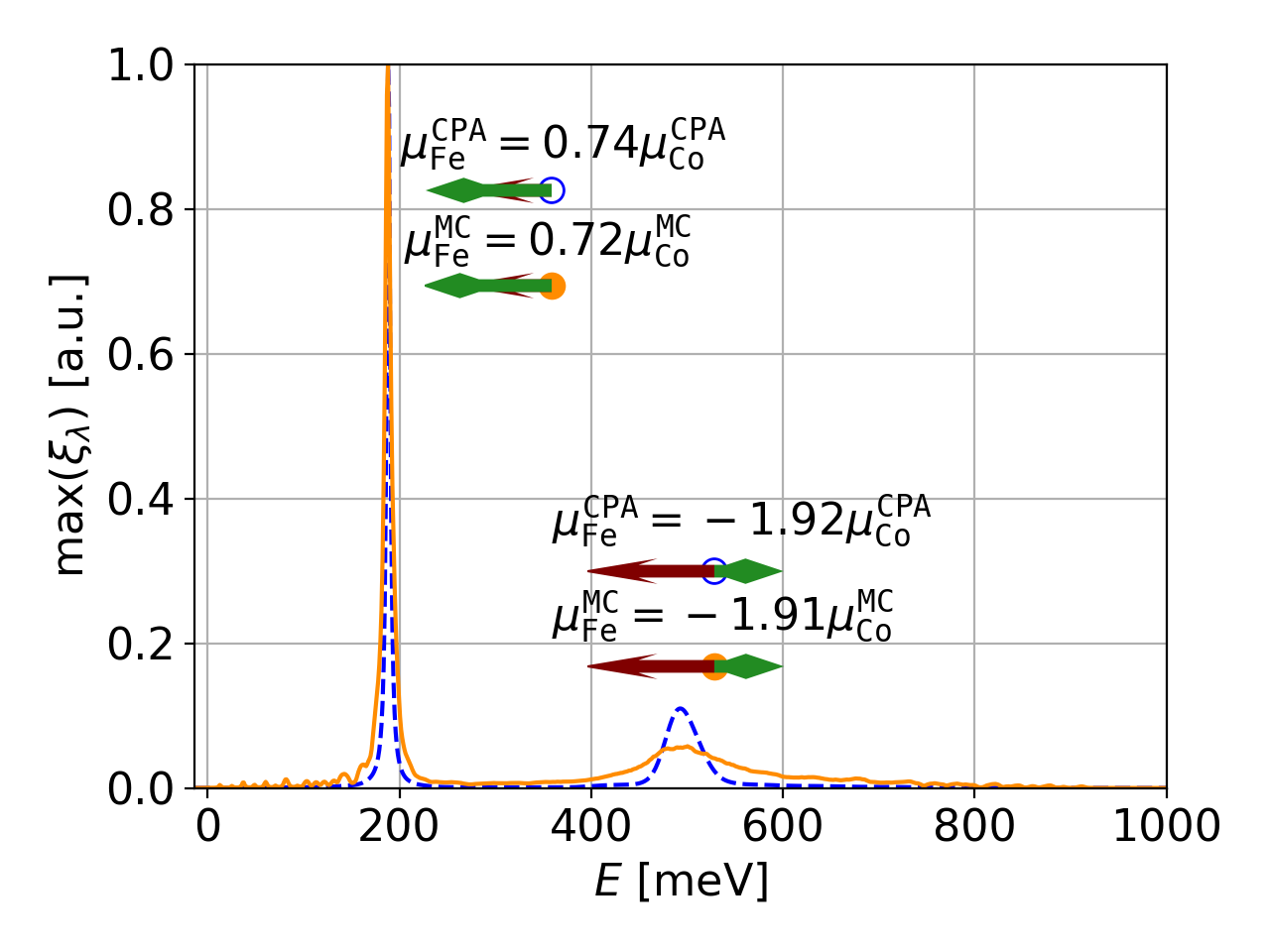}
  \includegraphics[width=0.45\textwidth]{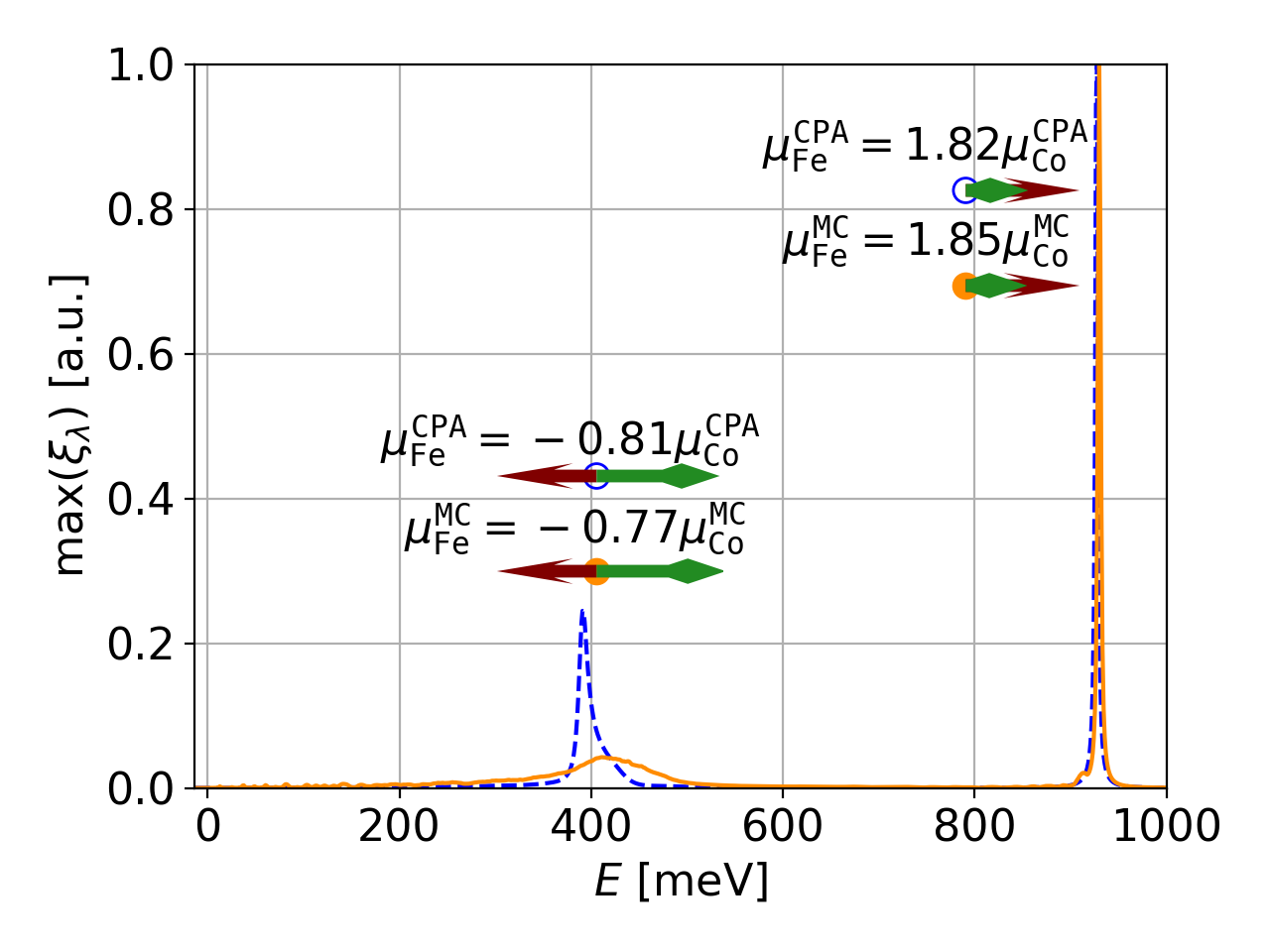}
  \includegraphics[width=0.45\textwidth]{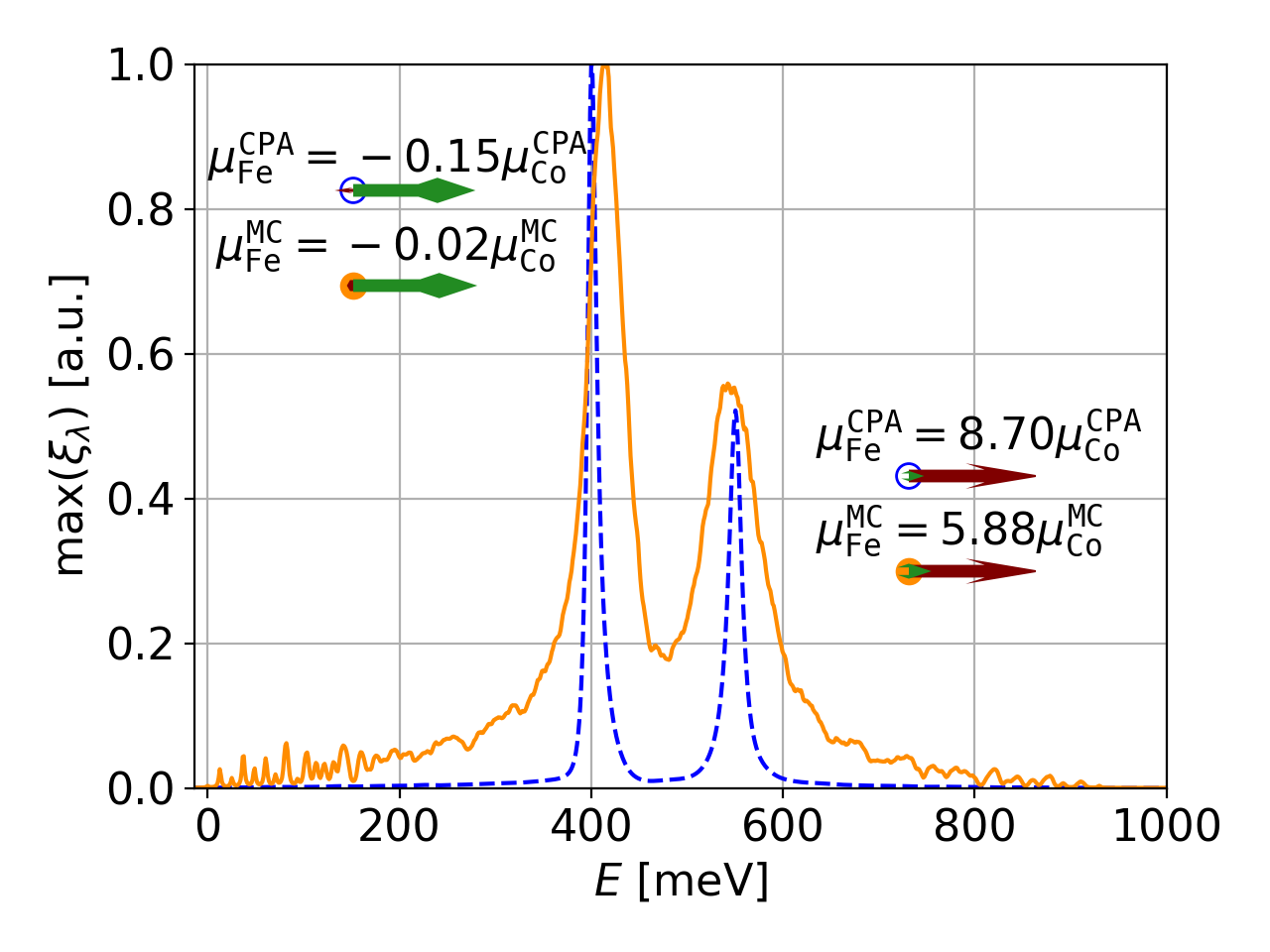}
  \caption{Maximal eigenvalue of the loss matrix at $\Gamma$
    (top left), $\frac{\Gamma\text{H}}{3}$ (top right), H
    (bottom left) and N (bottom right). All results are
    calculated at $T=$0\ K within the MC (orange solid line) and
    \ac{cpa} (blue dashed line). The spatial shape of the modes
    is drawn near the peaks. Starting from a blue empty circle
    (\ac{cpa}) or an orange filled circle (MC), the magnetic
    moments of iron are drawn as brown pointy vectors while the
    moments of cobalt are visualized as green edgeless vectors.}
  \label{fig_modes_G} 
\end{figure}

The red dots in figure \ref{fig_spectrum} mark the $k$-points for which the spatial shapes of the magnon modes will be analyzed in more detail now. 
In figure \ref{fig_modes_G} the eigenvalues and
eigenvectors of the loss matrix are shown at different symmetry points
in the \ac{bz} and $T=0$. The maximal eigenvalue at the $\Gamma$ point
for $T=0$ (figure \ref{fig_modes_G}, top left) exhibits the Goldstone
mode at zero energy, i.e. the mode where the moments of both
constituents are tilted by the same amount. 

As already mentioned, we explain the larger width of the MC peaks with
the influence of correlation effects between different sites, which
the \ac{cpa} cannot account for.  The appearance of smaller peaks at
regions where the \ac{cpa} susceptibility is effectively zero is a
further effect caused by non local effects. An interesting fact is
that at the N point (c.f. bottom right plot in figure
\ref{fig_modes_G}) essentially only one of the constituents precesses
for both the acoustic and the optic mode. Unfortunately, the direct measurement of the shape very challenging but recent studies suggest that it could be realized on the surface of a 2D magnet using atomic resolution inelastic scanning tunneling microscopy \cite{Hirjibehedin2006}.

Generally, we come to the conclusion that the CPA is able to
precisely give the magnetic properties of this iron cobalt alloy
apart from a systematic underestimation of the magnon damping. 

\subsection{Spatial form of the eigenmodes}
\begin{figure}
	\centering
	\includegraphics[width=\textwidth]{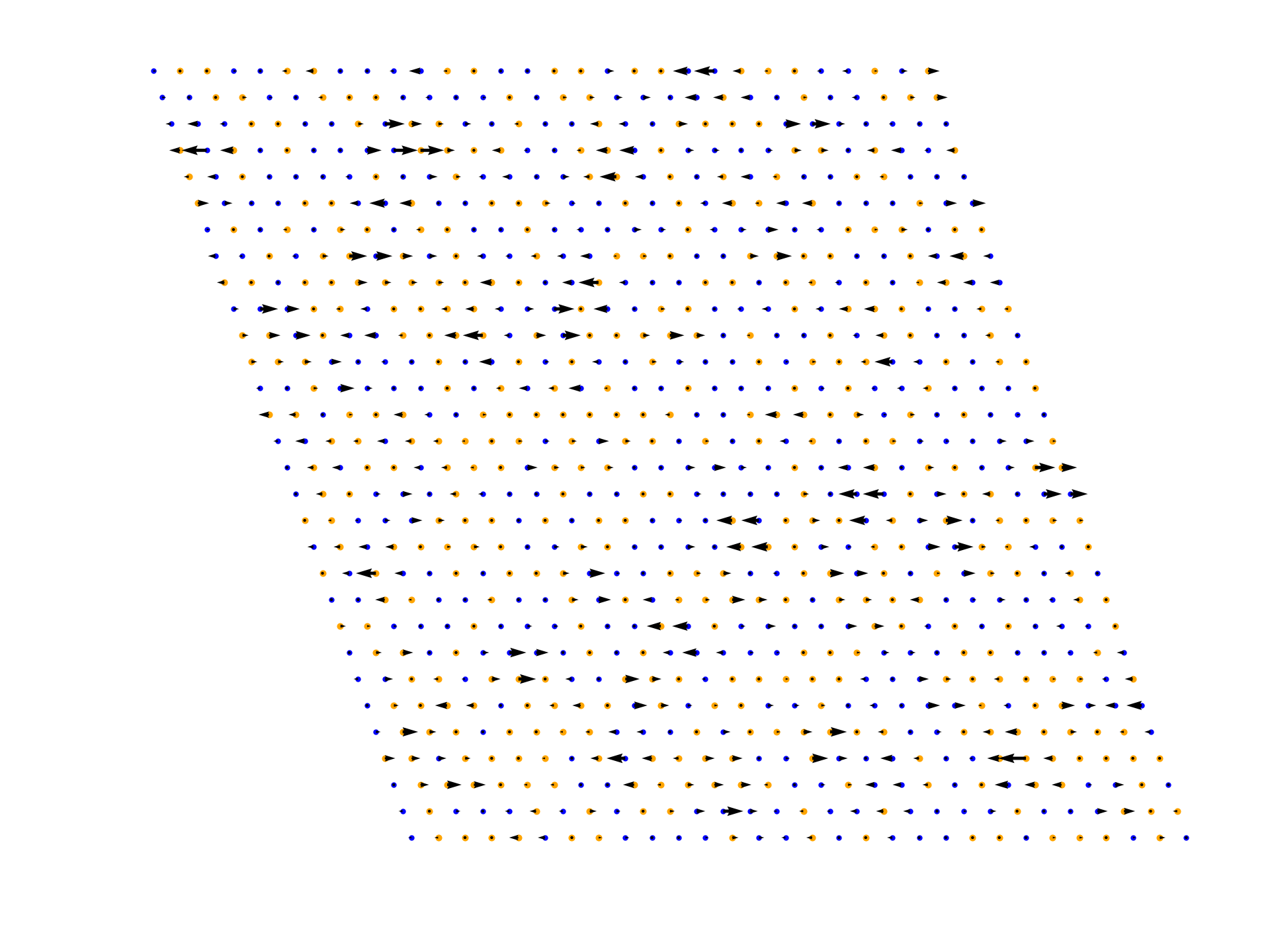}
	\caption{Real space mode of the susceptibility (corresponding
          to the highest eigenvalue) at $E=200\text{meV}$. The
          blue/orange dots represent iron/cobalt atoms.} 
	\label{fig_rsmode1}
\end{figure}

\begin{figure}
	\centering
	\includegraphics[width=0.45\textwidth]{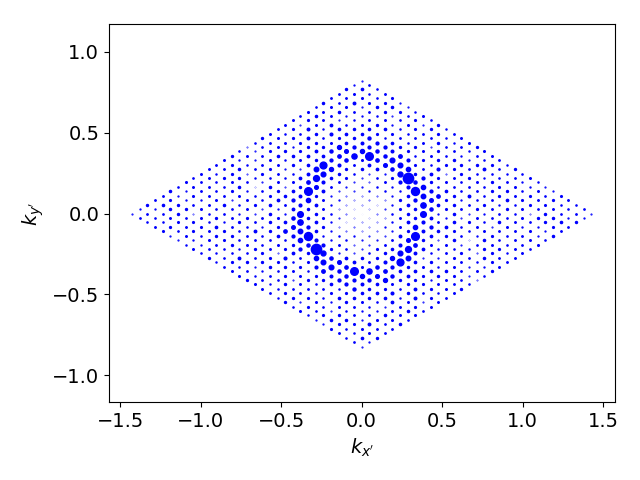}
	\includegraphics[width=0.45\textwidth]{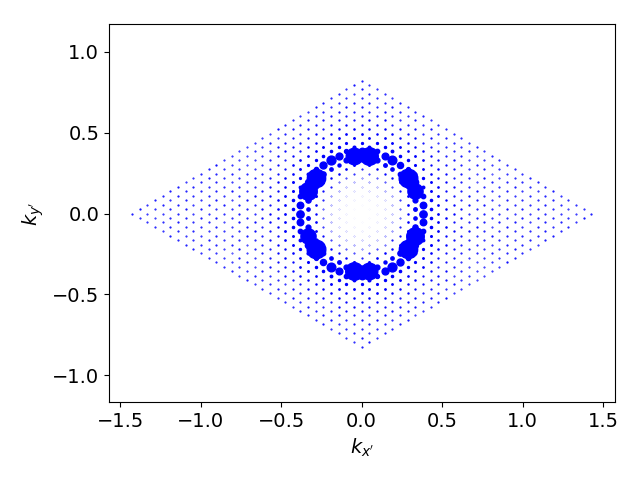}
    \caption{Fourier components of the eigenmode corresponding to
    	the highest eigenvalue $\xi_1$ (left) and the loss matrix
    	projected on plain wave states (right) for different
    	reciprocal lattice vectors. The plane shown in the figure correspond to the space spanned by the reciprocal lattice vectors $b_1=\left(0,\frac{2\pi}{a},\frac{2\pi}{a}\right)$ and $b_2=\left(\frac{2\pi}{a},0,\frac{2\pi}{a}\right)$.}
	\label{fig_rsanalysis1}
\end{figure}

Through the diagonalization of the susceptibility in real space
(equation \ref{eqn_susc}), the spatial eigenvectors at a certain
energy can be directly found through the loss matrix (equation
\ref{eqn_loss}). We did so calculating the real space susceptibility
of one specific random configuration in a 30$\times$30$\times$10 atom
supercell. The eigenmode corresponding to the highest eigenvalue of
the loss matrix at $E\approx200\text{meV}$ is depicted for this
specific random configuration in figure \ref{fig_rsmode1}. This figure
shows one plane of the supercell corresponding to
$\vb*{R}=\lambda_1\vb{a}_1+\lambda_2\vb{a}_2+5\vb{a}_3$ with the
primitive lattice vectors $\vb*{a}_i$,
$\lambda_1,\lambda_2\in\mathds{N}$ and
$\lambda_1,\lambda_2\leq30$. The orange dots correspond to \ch{Co}
atoms while the blue dots represent \ch{Fe} atoms. It is assumed that
in the magnetic ground state all magnetic moments are oriented
orthogonal to the plane depicted. The in-plane components of the
magnetic moments $\tilde{\mu}_i$ of this particular mode are represented by the arrows. Interestingly, we observe that the
mode shown in figure \ref{fig_rsmode1} features multiple small clusters of
precessing moments.
We further analyze the eigenmode by performing a Fourier analysis of the $\tilde{\mu}_i$
\begin{align}
	\tilde{\mu}_{\vb*{k}}=\frac{1}{N}\sum_{i=1}^{N}\text{e}^{\text{i}\vb*{k}\cdot\vb*{R}_i}\tilde{\mu}_i
\end{align}
with the positions of the atoms $\vb*{R}_i$ and the number of atoms $N=9000$ in our case.
The result is depicted in figure \ref{fig_rsanalysis1} for a plane in
the \ac{bz}. The dominant contributions arise from wavevectors lying
on an circle. In an ordered system, the circle would be the cross section of the constant energy surface in reciprocal space. 
Due to the disorder, Bloch waves cease to be eigenstates of the system (the corresponding peaks acquire finite widths) and the eigenstates pick up Fourier components outside of a single energy surface.
Finally, we further analyze the loss matrix by writing its spectral representation
\begin{align}
	\vb*{\mathcal{L}}_{ij}(E)=\sum_{\lambda}\xi_\lambda\ket{\tilde{\mu}_{i}^{\lambda}}\bra{\tilde{\mu}_{j}^{\lambda}}
\end{align}
with the eigenvalues $\xi_\lambda$ and eigenvectors
$\ket{\tilde{\mu}_{i}^{\lambda}}$. The projection of the loss matrix
to plain wave states is then given by 
\begin{align}
	\bra{\vb*{k}}\vb*{\mathcal{L}}(E)\ket{\vb*{k}}=\sum_\lambda\xi_\lambda(E)\abs{\tilde{\mu}_{\vb*{k}}^{\lambda}}^2
\end{align}
which represents a weighted sum of Fourier components. The plain wave projection for all eigenvalues larger than $\frac{\xi_1}{100}$,
with the highest eigenvalue $\xi_1$, is given in figure
\ref{fig_rsanalysis1}. Considering all the modes with significant eigenvalues at this energy recovers the picture of the constant energy surface similar to the one of an ordered system. While a single eigenmode is clearly different from a Bloch wave, the dynamics of the entire system resembles the one of the ordered system. In a sense, the Fourier transformation recovers the self-averaging property of the spin dynamics in disordered magnets.

\subsection{Eigenmodes at finite temperatures}
\begin{figure}
  \centering
  \includegraphics[width=0.45\textwidth]{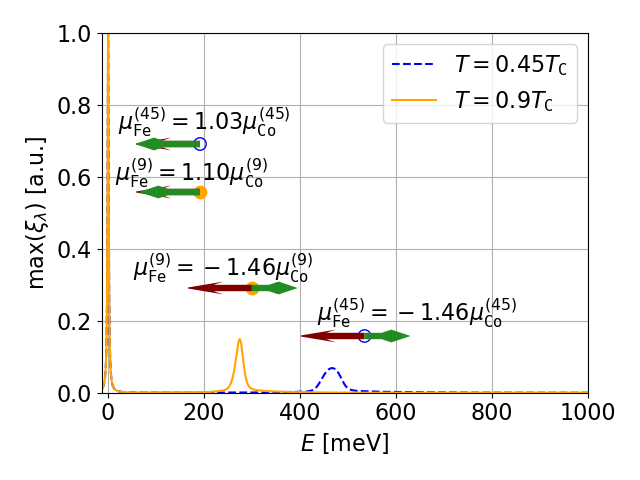}
  \includegraphics[width=0.45\textwidth]{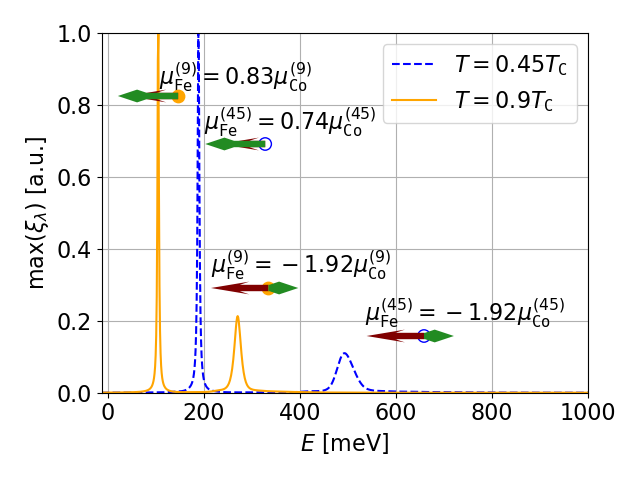}
  \includegraphics[width=0.45\textwidth]{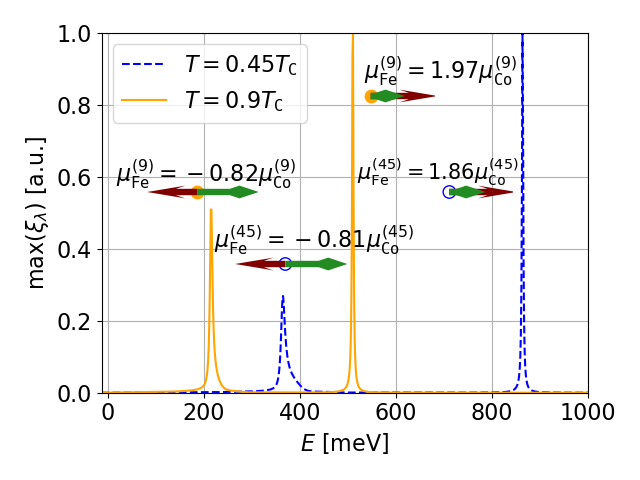}
  \includegraphics[width=0.45\textwidth]{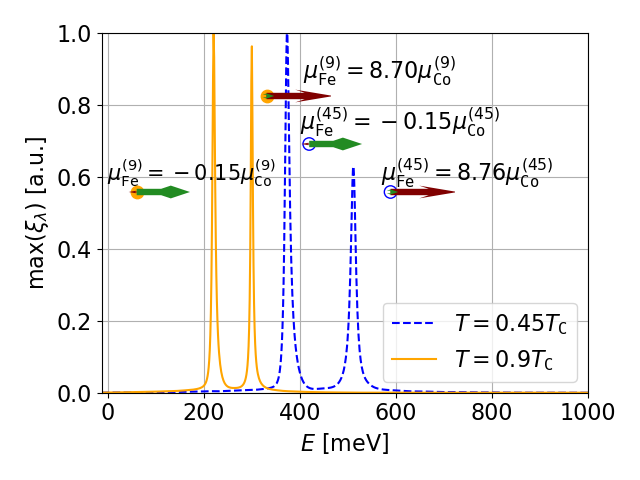}
  \caption{Maximal eigenvalue of the loss matrix at $\Gamma$
    (top left), $\frac{\Gamma\text{H}}{3}$ (top right), H
    (bottom left) and N (bottom right). All results are
    calculated at $T=0.45T_\text{C}$ and $T=0.9T_\text{C}$
    within the CPA. The spatial shape of the modes is drawn near
    the peaks. Starting from a blue empty circle
    ($T=0.45T_{\text{C}}$) or an orange filled circle
    ($T=0.9T_{\text{C}}$), the magnetic moments of iron are
    drawn as brown pointy vectors while the moments of cobalt
    are visualized as green edgeless vectors.}
  \label{fig_modes_Gt} 
\end{figure}

Next, we investigate the change of the eigenmodes with temperature. It
turned out that our realization of the RPA in combination with the MC
calculations is computationally too expensive for the system size we
are considering here. Therefore, we restrict this discussion to the
\ac{cpa}+\ac{rpa} results and stress again the agreement of \ac{cpa} and \ac{mc} shown in
the previous section especially when it comes to the spatial form of
the modes. We recall that the widely used \ac{rpa} \cite{Tang2006,Bouzerar2002,Matsubara1973} cannot account for the temperature broadening of the magnon modes \cite{Paischer2021} such that we must restrict ourselves to the analysis of the impact of the temperature on the shapes of magnetic modes, cf. figure \ref{fig_modes_Gt}. 
Our calculations
suggest that the spatial shape of the modes is independent of
temperature. 

\section{Summary}
We provided a thorough analysis of the magnonic modes in a disordered iron
cobalt alloy using two complementary numerical schemes. The \ac{mc}
and the \ac{cpa} give basically the same magnonic properties apart from
the disorder induced width of the peaks in the magnonic spectrum. We explain this discrepancy with non-local
effects which the single site \ac{cpa} cannot account
for. Interestingly, we found that the acoustic and optic mode at the N
point features the precession of only one of the constituents'
magnetic moments. 
The eigenspectrum analysis of the loss matrix in real space reveals that the eigenmodes
involve many small clusters
of precessing magnetic moments. We are convinced that such real space effects can be used to effectively excite only defining parts of the magnonic crystal and thus gain additional precise control over the spin dynamics of the system.
Finally, it turns out that the spatial shape of the modes is
basically invariant with respect to temperature.
\bibliography{./Quellen}

\end{document}